\documentclass[aps,prb,twocolumn,reprint,superscriptaddress,notitlepage]{revtex4-1}
\usepackage{graphicx}
\usepackage{bm}
\usepackage{amsmath}
\usepackage{amssymb}
\usepackage{amsfonts}
\usepackage{euscript}
\usepackage{verbatim}
\usepackage{setspace}
\usepackage{xcolor}
\usepackage{amsfonts}
\usepackage{braket}
\usepackage{qcircuit}
\usepackage{verbatim}

\usepackage[normalem]{ulem}
\usepackage{hyperref}

\usepackage{cleveref}

\newcommand{\be}{\begin{equation}}
\newcommand{\ee}{\end{equation}}
\newcommand{\bea}{\begin{eqnarray}}
\newcommand{\eea}{\end{eqnarray}}

\newcommand{\Tr}[1]{\mathrm{Tr} #1}
\newcommand{\red}[1]{{\color{red}{#1}}}

\newcommand{\emanuele}[1]{{\color{purple}{#1}}}

\newtheorem{thm}{Theorem}
\crefname{thm}{Theorem}{Theorems}

\newcommand{\mycite}[1]{\onlinecite{#1}}

\begin{document}
\title{Quantum and classical Floquet prethermalization}

\author{Wen Wei Ho}
\affiliation{Department of Physics, National University of Singapore, Singapore 117542}

\author{Takashi Mori}
\affiliation{RIKEN Center for Emergent Matter Science (CEMS), Wako 351-0198, Japan}

\author{Dmitry A. Abanin}
\affiliation{Department of Theoretical Physics, University of Geneva, Quai Ernest-Ansermet 30, 1205 Geneva, Switzerland}

\author{Emanuele G. Dalla Torre}
\affiliation{Department of Physics, Bar-Ilan University, Ramat Gan 5290002, Israel}
\affiliation{Center for Quantum Entanglement Science and Technology,
Bar-Ilan University, Ramat Gan 5290002, Israel}

\begin{abstract}


Time-periodic (Floquet) driving is a powerful way to control the dynamics of complex systems, which can be used to induce a plethora of new physical phenomena. However, when applied to  many-body systems, Floquet driving can also cause heating, and lead to a featureless infinite-temperature state, hindering most useful applications. It is therefore important to find mechanisms to suppress such effects. 
Floquet prethermalization refers to the phenomenon where many-body systems subject to a high-frequency periodic drive avoid heating for very long times, instead tending to transient states that can host  interesting physics.
Its key signature is a strong parametric suppression of the heating rate as a function of the driving frequency. Here, we review our present understanding of this phenomenon in both quantum and classical systems, and across various models and methods. In particular, we present rigorous theorems underpinning Floquet prethermalization in quantum spin and fermionic lattice systems, extensions to systems with degrees of freedom that have unbounded local dimension. Further, we briefly describe applications to novel nonequilibrium phases of matter, and recent experiments probing prethermalization with quantum simulators. We close by describing the frontiers of Floquet prethermalization beyond strictly time-periodic drives, including time-quasiperiodic driving and long-lived quasi-conserved quantities enabled by large separation of energy scales.

\end{abstract}

\maketitle

\section{Introduction}

Quantum systems with parameters that are periodically varied in time, called Floquet systems, have recently attracted much interest. First,  time-periodic electromagnetic fields 
allow for the control of 
effective Hamiltonians in systems of cold atoms in optical lattices~\cite{eckardt2017colloquium,weitenberg2021tailoring}, and in quantum materials~\cite{oka2019floquet}. In particular, periodic driving can be used to engineer band structures, with applications including creating artificial gauge fields and topological Bloch bands~\cite{rudner2020band,weitenberg2021tailoring} -- a research direction often referred to as  {\it Floquet engineering}. Second, periodic driving can induce intrinsically non-equilibrium physical phenomena with no static counterpart. 
A celebrated example is the Thouless pump~\cite{thouless1983}, which features quantized transport. More recently, {\it Floquet phases of matter}, characterized by spatiotemporal responses not exhibited by  phases of matter at equilibrium, such as the discrete time-crystals~\cite{sondhi2017} and the anomalous Floquet Anderson insulator (AFAI)~\cite{rudner2020band}, have been uncovered.

However, there is an inherent difficulty in employing periodic driving for  systems with many degrees of freedom, related to the fact that driving is expected to induce heating in such systems. According to the second law of thermodynamics, the entropy of isolated many-body systems always increases up to its maximum value at equilibrium, and is constrained only by global conservation laws. In the context of many-body systems which are being driven, since energy conservation is broken,  such systems are   expected to continuously  heat up over time (that is, in the absence of dissipation).  One may then anticipate  that at late times  these  systems tend to a ``boring'', featureless infinite temperature ensemble, wherein all microstates occur with equal probability\footnote{More precisely, for {\it quantum} many-body systems, this should be understood at the level of local subsystems. That is, the entropy in question is the von Neumann entropy associated with the reduced density matrix of a small region in space, and the expectation is that such density matrices tend to an infinite-temperature Gibbs state under a drive.}. This effect  poses a major challenge for using  periodic drives to create non-trivial couplings via Floquet engineering and to realize novel Floquet phases in many-body systems.

It is therefore important to devise ways in which   heating in Floquet systems  can be eliminated or strongly  suppressed, such that the heating rate can be made arbitrarily small by an easily tunable external parameter. Among the possible strategies that have been studied recently, two common ones are (i) to drive integrable systems in a way that retains their integrability, which prevents thermalization by  preserving a large number of conserved quantities \cite{russomanno12periodic,gritsev2017integrable,ishii2018heating}, 
and (ii) to impose strong disorder, which gives rise to robust emergent integrability  and suppresses heating by trapping the system in non-thermal, many-body localized states \cite{ponte15periodically,ponte2015many,lazarides15fate}. 

In this review, we consider another general mechanism to suppress heating in  many-body Floquet systems, called {\it Floquet prethermalization}. This method relies on the application of a periodic drive at a large driving frequency, and does not require integrability or disorder (and can, in fact, be combined with either). In the case of simple quantum systems, such as a single atom or a molecule with few levels, high-frequency drives are routinely studied by moving into a rotating frame. In the new frame the effect of the fast drive can be `integrated out', leading to an effective static Hamiltonian, known as the Floquet Hamiltonian, that governs dynamics. 
This approach, commonly used in quantum optics,  can be formally expressed as a controlled Magnus expansion, with the inverse of the driving frequency playing the role of a small parameter~\cite{scully1997quantum,bukov2015universal}. A similar approach  has been applied for many decades  in the context of nuclear magnetic resonance (NMR)  to describe the behavior of nuclear spins under periodic electromagnetic field pulses \cite{farrar2012pulse}.

\begin{figure}[t]
\centering
\includegraphics[scale=0.4]{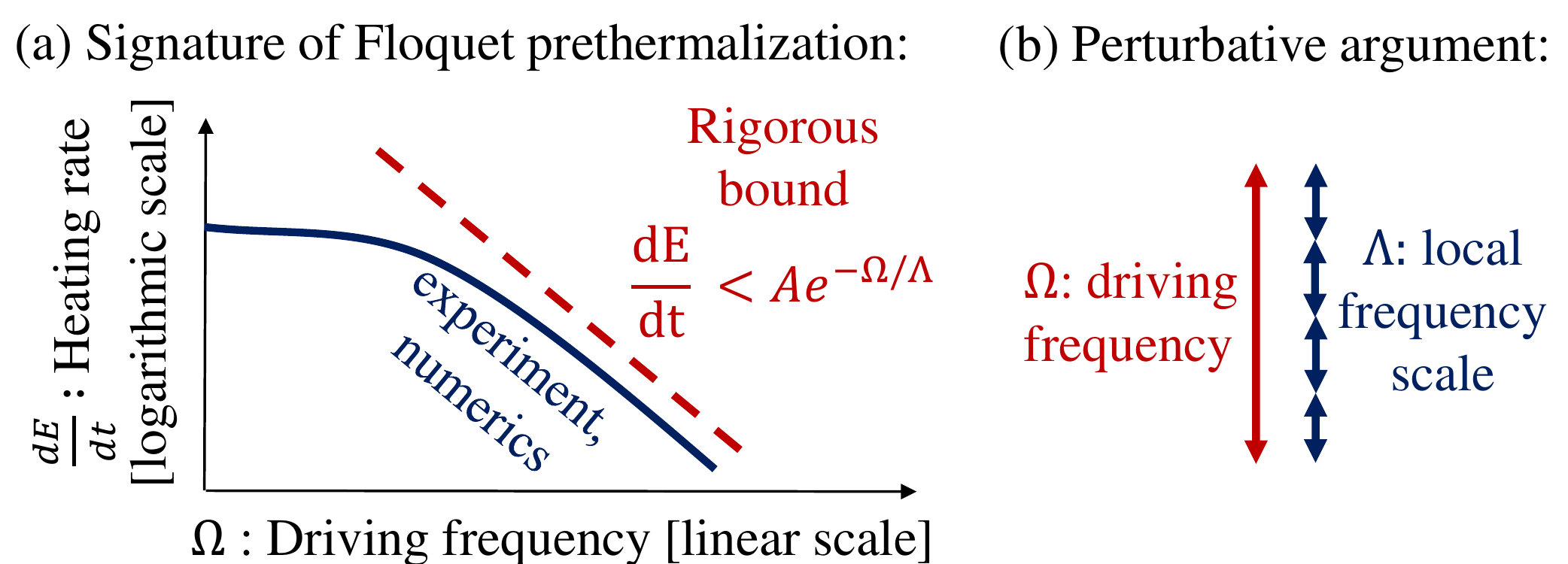}
\caption{(a) Schematic description of the rigorous bounds on Floquet prethermalization: in many physically relevant scenarios, the heating rate at large driving frequencies is exponentially suppressed in $\Omega/\Lambda$, where $\Lambda$ is a local frequency scale.
(b) This suppression can be understood using a perturbative argument (see text):
it takes $\sim \Omega/\Lambda$ local reconfigurations of the system, each changing the system's energy by $\Lambda$ to absorb a large quantum of energy $\Omega$ from the drive. Such a process is exponentially suppressed in $\Omega/\Lambda$. 
} 
\label{fig:schematic1}
\end{figure}

Upon na\"ively carrying over similar manipulations for strongly interacting systems with a large number of degrees of freedom, one might conclude that at sufficiently high frequencies, 
heating in such systems is absent,  owing to a  conservation of `energy' of the Floquet Hamiltonian. 
In particular, thermodynamic concepts, like the notion of equilibrium states at an effective temperature, should then be applicable, provided one observes the system in the appropriate   rotating  frame. However, a more careful consideration shows that this conclusion cannot in general be correct: 
 the extensivity of the many-body bandwidth and the denseness of the spectrum of a thermodynamically large quantum system imply that any drive with energy quanta that carry a finite frequency will inevitably lead to energy absorption, due to the proliferation of resonant transitions between   many-body eigenstates\footnote{See Ref.~\protect\mycite{seetharam2018absence} for a study of the crossover between small and large systems.}.
Mathematically, this is reflected in the fact that the Magnus expansion has a finite radius of convergence which depends on the ratio of the many-body bandwidth of the system, which is an extensive quantity, and the driving frequency \footnote{See ref.~\protect\mycite{casas2007sufficient,d14long} and references therein.}. Thus, the driving frequency needs to scale at least as fast as the system size in order for the  high-frequency expansion to converge. 
This is arguably not a very physical scenario,  especially in the limit  of thermodynamically  large systems.
Despite such a fundamental obstacle for avoiding heating in many-body systems,  recent works have shown that the {\it rate} of heating can nevertheless be strongly suppressed at large driving frequencies, with a corresponding static Hamiltonian which {\it is} approximately conserved for very long times  \cite{abanin15exponentially,kuwahara16floquet, mori2016rigorous,abanin2017rigorous,abanin17effective}. 
Importantly, this Hamiltonian has a local structure (i.e., it has interactions decaying sufficiently fast with spatial distance), allowing for a meaningful interpretation of it as an effective energy operator and associated with an equilibrium-like Boltzmann distribution. Often, this operator can be constructed by a  {\it truncation} of a high-frequency expansion at an optimal order. 
Concretely, for lattice quantum models with sufficiently local interactions (spins or fermions), the heating rate can be rigorously bounded by an exponential function of the ratio between the driving frequency $\Omega$ and a typical local frequency scale of the system $\Lambda$, which importantly is not extensive, see Fig.~\ref{fig:schematic1} (a). 
In other cases, like those of bosons on a lattice, although a rigorous bound cannot be derived, statistical arguments can be used to derive an exponential relation between the heating rate and $\Omega$. In both situations, the phenomenon of parameterically slow heating is often referred to as Floquet prethermalization, where the ``pre-'' prefix indicates that 
the system has not reached its infinitely long-time behavior yet; instead, 
it achieves a quasi-equilibrium state often described by 
a {\it thermal} ensemble with respect to the almost-conserved energy operator, with parameters (like temperature) that change slowly in time.

Intuitively, the mechanism of Floquet prethermalization can be  understood using a linear response argument\cite{abanin15exponentially}.
If we consider a time-periodic drive with a small amplitude $g$ and large frequency $\Omega$, the dominant contribution to heating is given by processes wherein one quantum of excitation from the drive is transferred to the system. According to Fermi's Golden Rule, the rate of this process is proportional to   $g^2$  times the square of the transition matrix elements between energy eigenstates separated by energy $\Omega$. For a many-body system one can estimate this matrix elements as follows:  Connecting two states that are separated by an energy $\Omega$ requires a number $n$ of local rearrangements (e.g., spin flips), where $n \propto \Omega/\Lambda$. Accordingly, the matrix element connecting these two states is exponentially small in $n$, leading to an exponential suppression of the heating rate.
However, as we will explain below in detail, Floquet prethermalization is not limited  to this perturbative regime and extends to drives whose amplitudes can  be comparable to those of local couplings.


The goal of this review is to introduce the concept of Floquet prethermalization, in a compendious, yet physically accessible fashion.  
A bulk of this review (Sec.~\ref{sec:rigorous}) focuses on quantum systems of spins or fermions on a lattice, where mathematical bounds on heating were first derived. Emphasis is placed on explaining the degree of rigor in each result, and on the relation between existing approaches. 
In Sec.~\ref{sec:quantum2classical} we discuss extensions   these rigorous results to classical systems, putting a special emphasis on differences between classical and quantum cases. In Sec.~\ref{sec:statistical} we move to many-body systems where rigorous methods do not apply and introduce the idea of statistical Floquet prethermalization.  In Sec.~\ref{sec:extensions} we present several theoretical extensions and applications of Floquet prethermalization. Sec.~\ref{sec:experiments} addresses recent experiments aimed at demonstrating Floquet prethermalization in synthetic systems and materials. Sec.~\ref{sec:summary} concludes the review and describes future research directions.

\def \hat {{}} 

\section{Rigorous results for quantum spin systems}

\label{sec:rigorous}

Consider a quantum system  described by a time dependent Hamiltonian $H(t)$ with time period $T=2\pi/\Omega$, such that $H(t+T)=H(t)$. Its stroboscopic time evolution
is conveniently described by the Floquet Hamiltonian $H_{\rm F}$
defined\footnote{Note though the Floquet Hamiltonian is not unique: there is a choice of branch cut involved in taking the logarithm of $U(0,T)$).}  by $U(0,T)\equiv\exp(-i H_{\rm F} T)$, where the Floquet unitary $U(0,T)$ is the evolution operator over one time period. Thanks to     linearity of the Schr\"odinger equation,   stroboscopic evolution is simply given by multiple applications of this operator:
\begin{align}
U(0,nT) = \left[U(0,T)\right]^n = \exp(-i n H_{\rm F} T),
\end{align}
where $n$ is an integer. (This identity can also be understood as arising from the Floquet theorem.)
Thus, we may view time evolution over a time $nT$ as being equivalent to  time evolution described by the time-independent Hamiltonian $H_{\rm F}$. 
One may be tempted to conclude that due to the exact conservation of $H_F$ at stroboscopic times, 
the system has a conserved energy,
so that heating does not occur. However, the operator $H_{\rm F}$   is in general a  complicated function of the original time-dependent Hamiltonian $H(t)$ and, for many-body systems, it is not expected to be expressible as a sum of local terms. The presence of highly nonlocal terms hinders
an interpretation of $H_{\rm F}$ as a physically meaningful total energy operator for a many-body system,
and consequently prevents the application of thermodynamic considerations, such as the definition of a temperature with respect to $H_{\rm F}$. 


In what follows, we will demonstrate two rigorous approaches aimed at finding an effective Hamiltonian $H_{\rm eff}$ which possesses a local structure, and hence {\it does} play the role of the total energy of the system, which in turn can be used to bound the system's heating rate \cite{kuwahara16floquet, mori2016rigorous,abanin2017rigorous,abanin17effective}. To explain these results, it will be necessary to first introduce some technical tools needed to 
quantify 
the strength, or amplitude of a drive. For a many-body system, simply using the operator norm of the driving term would result in an estimate of its amplitude as being divergent with system size (owing to its many-body nature), which is typically not very useful. To resolve this problem,  we consider  a particular class of systems, many-body systems of quantum spins or fermions whose  local state space is bounded, and require that their interactions are local, either decaying sufficiently fast with geometrical distance or acting only on finite subsets of degrees of freedom. 
We note that this assumption covers a large class of physically relevant Hamiltonians: for example, the celebrated transverse field Ising model and Heisenberg model both only have nearest-neighbor interactions. 
For this class of systems, it is possible to define a local norm of the Hamiltonian, which, roughly speaking, amounts to measuring the strength of all interaction terms that affect a given site.
This amplitude, denoted by $\Lambda$, is an intensive quantity, and has a clear physical interpretation as the energy cost of performing a local rearrangement of the system (e.g., flipping a spin), which does not to scale with system size.
For example, in spin systems, the scale $\Lambda$ can be determined by Zeeman fields or (local) Ising/Heisenberg couplings, and in itinerant fermionic systems by hopping and local on-site interactions. 

To formally define $\Lambda$, we consider a quantum spin (or fermion) system on a $d$-dimensional regular lattice.
Each lattice site is labeled by $i=1,2,\dots,N$ with $N$ being the total number of lattice sites.
We assume that the Hamiltonian is expressible as
\begin{align}
    H(t)=\sum_{X:|X|\leq k}h_X(t),
    \label{eq:k-local}
\end{align}
where $X$ is a subset of the sites of the lattice and $h_X(t)$ is an operator  acting nontrivially only on region $X$.
The condition   $|X|\leq k$ means that $X$ contains at most $k$ different sites,  
 i.e., the Hamiltonian is such that it has at most $k$-site interactions.
For this reason, eq.~(\ref{eq:k-local}) is sometimes referred to as a $k$-local Hamiltonian.
A useful notion of local norm $\Lambda$ of a time-periodic Hamiltonian $H(t)$ can then be defined as $\Lambda=\max_{t\in[0,T]}\Lambda(t)$, where the instantaneous bound $\Lambda(t)$ is defined as
\begin{align}
    \Lambda(t)=\max_{i\in\{1,2,\dots,N\}}\sum_{X:|X|\leq k, i\in X}\|h_X(t)\|,
    \label{eq:local_norm}
\end{align}
where 
$\|\cdot\|$ denotes the operator norm and the sum runs over all subsets of sites that include the site $i$.
Intuitively, $\Lambda(t)$ measures the largest amplitude of a transition that can affect any given site due to $H(t)$. Slightly different definitions of $\Lambda$ are used in the literature, but they are similar in spirit to \cref{eq:local_norm}. For example, it is possible to define $\Lambda$ to include nonlocal terms, as long as the norm of the sum of all terms $h_X(t)$ with $|X|=S$ decays exponentially in $S$, see ref.~\mycite{abanin2017rigorous}. 

With this set-up, we can now summarize the  rigorous results regarding slow heating of many-body systems with finite local norm $\Lambda$. 

\subsection{Summary of rigorous results}

\red{}

\begin{thm}\label{thm:quasi}
Suppose a $k$-local Hamiltonian $H(t)=H(t+T)$  is time-periodic with period $T=2\pi/\Omega$ and has a local norm $\Lambda$. We denote by $H_0=(1/T)\int_0^TH(t)dt$ the static part of the Hamiltonian and by $\ket{\psi(t)}$ a solution of the Schr\"odinger equation $i\partial_t\ket{\psi(t)}=H(t)\ket{\psi(t)}$.
Then, we can find an effective Hamiltonian $H_\mathrm{eff}$ that satisfies the following properties:
\begin{align}
\frac{1}{N}\|H_\mathrm{eff}-H_0\|=\mathcal{O}\left(\frac{\Lambda}{\Omega}\right)
\label{eq:quasi1}
\end{align}
and
\begin{align}
\frac{1}{N}|\braket{\psi(t)|H_\mathrm{eff}|\psi(t)}-\braket{\psi(0)|H_\mathrm{eff}|\psi(0)}|\leq e^{-\mathcal{O}(\Omega/\Lambda)}\Lambda^2 t
\label{eq:quasi2}
\end{align}
for an arbitrary initial state $\ket{\psi(0)}$ at   times $t=nT$ with $n$ a positive integer. Here, $\| \cdot \|$ is the standard operator norm. 
\end{thm}

Eq.~\ref{eq:quasi1} means that $H_\mathrm{eff}$ is close to the static Hamiltonian $H_0$, and
\cref{eq:quasi2} tells us that $H_{\rm eff}$ is approximately conserved over an exponentially long time interval. Combining these two equations, one obtains that, under the conditions of \cref{thm:quasi}, the heating is exponentially slow in the following sense:
\begin{align}
    \frac{1}{N}\left|\braket{\psi(t)|H_0|\psi(t)}-\braket{\psi(0)|H_0|\psi(0)}\right|\leq e^{-\mathcal{O}(\Omega/\Lambda)}\Lambda^2 t+\delta,
    \label{eq:quasi3}
\end{align}
where $\delta=\mathcal{O}(\Lambda/\Omega)>0$ is a small constant, independent of $N$ and $t$. Eq.~\ref{eq:quasi3} indicates that $H_{\rm eff}$ is a quasi-conserved quantity: it is almost conserved 
for times shorter than an exponentially long heating time $\tau\sim \Lambda^{-1}e^{\mathcal{O}(\Omega/\Lambda)}$.
If there is no other (quasi-)conserved quantity, it is expected that the system relaxes locally to a prethermalized state described by the Gibbs state for the effective Hamiltonian:
\begin{align}
\rho_\mathrm{pre}\propto {e^{-\beta_\mathrm{eff}H_\mathrm{eff}}} ,
\label{eq:boltzmann}
\end{align}
where $\beta_\mathrm{eff}$ is the effective inverse temperature determined by the condition $\Tr[H_\mathrm{eff}\rho_\mathrm{pre}]=\braket{\psi(t)|H_\mathrm{eff}|\psi(t)}\approx\braket{\psi(0)|H_\mathrm{eff}|\psi(0)}$ for $t\ll\tau$.
In other words, a quasi-stationary state observed in the Floquet prethermalization is described by equilibrium statistical mechanics for the effective Hamiltonian~\citep{mori2017thermalization}.

Up to here, we have not assumed that the interactions are short-ranged i.e.~that they decay 
exponentially with the geometrical distance between two sites.  
Indeed, \cref{thm:quasi} implies that Floquet prethermalization can occur even in long-range interacting systems, as long as $H(t)$ is $k$-local (i.e., affects at most $k$ sites, no matter how far they are spatially separated) and has a finite local norm.
If we further require that the Hamiltonian is sufficiently spatially local so that the speed of information propagation is bounded to be within an algebraic light-cone -- which in systems with local interactions is guaranteed by the Lieb-Robinson bounds~\cite{lieb1972finite} -- then a stronger result can be proven. In this case, $H_\mathrm{eff}$ is not only a quasi-conserved quantity, but also {\it generates}  approximate time evolution of local observables, as shown by the following theorem.
\begin{thm}\label{thm:dyn}
Suppose a $d$-dimensional quantum spin system described by a $k$-local and short-ranged Hamiltonian that is time-periodic $H(t)=H(t+T)$ with $T=2\pi/\Omega$ and has a local norm $\Lambda$.
Denote by $\ket{\psi(t)}$ a solution of the Schr\"odinger equation $i\partial_t\ket{\psi(t)}=H(t)\ket{\psi(t)}$.
For any local observable $A$, we have
\begin{align}
    \left|\braket{\psi(t)|A|\psi(t)}-\braket{\psi(0)|e^{iH_\mathrm{eff}t}Ae^{-iH_\mathrm{eff}t}|\psi(0)}\right|\nonumber \\
    \leq C(\Lambda t+C')^{d+1}e^{-\mathcal{O}(\Omega/\Lambda)}
\end{align}
for any initial state $\ket{\psi(0)}$ at any stroboscopic time $t=nT$ with $n$ a positive integer, where $H_\mathrm{eff}$ is the effective Hamiltonian found in \cref{thm:quasi} and $C$, $C'$ are positive constants depending on the choice of the operator $A$.
\end{thm}

\cref{thm:quasi,thm:dyn} have been proved in two different ways: the Floquet-Magnus expansion~\citep{kuwahara16floquet,mori2016rigorous} and the renormalization method~\citep{abanin2017rigorous,abanin17effective}.
An explicit expression of the effective Hamiltonian $H_\mathrm{eff}$ is also obtained within each method.
In the following subsections, we review these two methods.

\subsection{Rigorous theory based on the Floquet-Magnus expansion}
\label{sec:FM}

The first method \cite{kuwahara16floquet, mori2016rigorous} relies on the Floquet-Magnus (FM) expansion~\cite{bcor}, which attempts to formally construct the Floquet Hamiltonian as a perturbative series in the driving period $T$:
\begin{align}
    \hat{H}_\mathrm{F}=\sum_{l=0}^\infty\hat{H}_lT^l.
    \label{eq:FM}
\end{align}
We can obtain the explicit form of $H_l$ by comparing two expressions of the exact time evolution operator, order by order:
\begin{align}
    U(0,T)=&1-i\int_0^Tdt_1\, H(t_1)\nonumber \\
    &-\int_0^Tdt_1\int_0^{t_1}dt_2 \, H(t_1)H(t_2)+\dots
\end{align}
and 
\begin{align}
    U(0,T)=e^{-iH_\mathrm{F}T}=1-iH_FT-\frac{T^2}{2}H_F^2+\dots
\end{align}
with \cref{eq:FM}.
For the first two terms, we obtain $H_1=\frac{1}{T}\int_0^T H(t)dt$ and
$H_2 = \frac{1}{2iT^2}\int_0^T dt_1 \int_0^{t_1} dt_2 [H(t_1),H(t_2)]$.
Note that $H_2$ is written  using a commutator  of Hamiltonians at two different times, $[H(t_1),H(t_2)]$.
By considering higher order terms, one can show that $\hat{H}_l$ corresponds to multiple nested commutators of the form $[\hat{H}(t_1),[\hat{H}(t_2),\dots,[\hat{H}(t_l),\hat{H}(t_{l+1})]\dots]]$ \footnote{see Supplemental Material of ref.~\protect\mycite{mori2022heating}.}.
%
Numerical calculations have demonstrated that  for a fixed period $T$, the radius of convergence of the FM expansion shrinks to zero as the system size $N$ grows to infinity\cite{d14long}. Nevertheless, in analogy to other asymptotic expansions, one can speculate that its first few terms can be used to describe the dynamics of the system at intermediate time scales, as will be shown below.

\begin{figure}[t]
    \centering
    \includegraphics[scale=0.5]{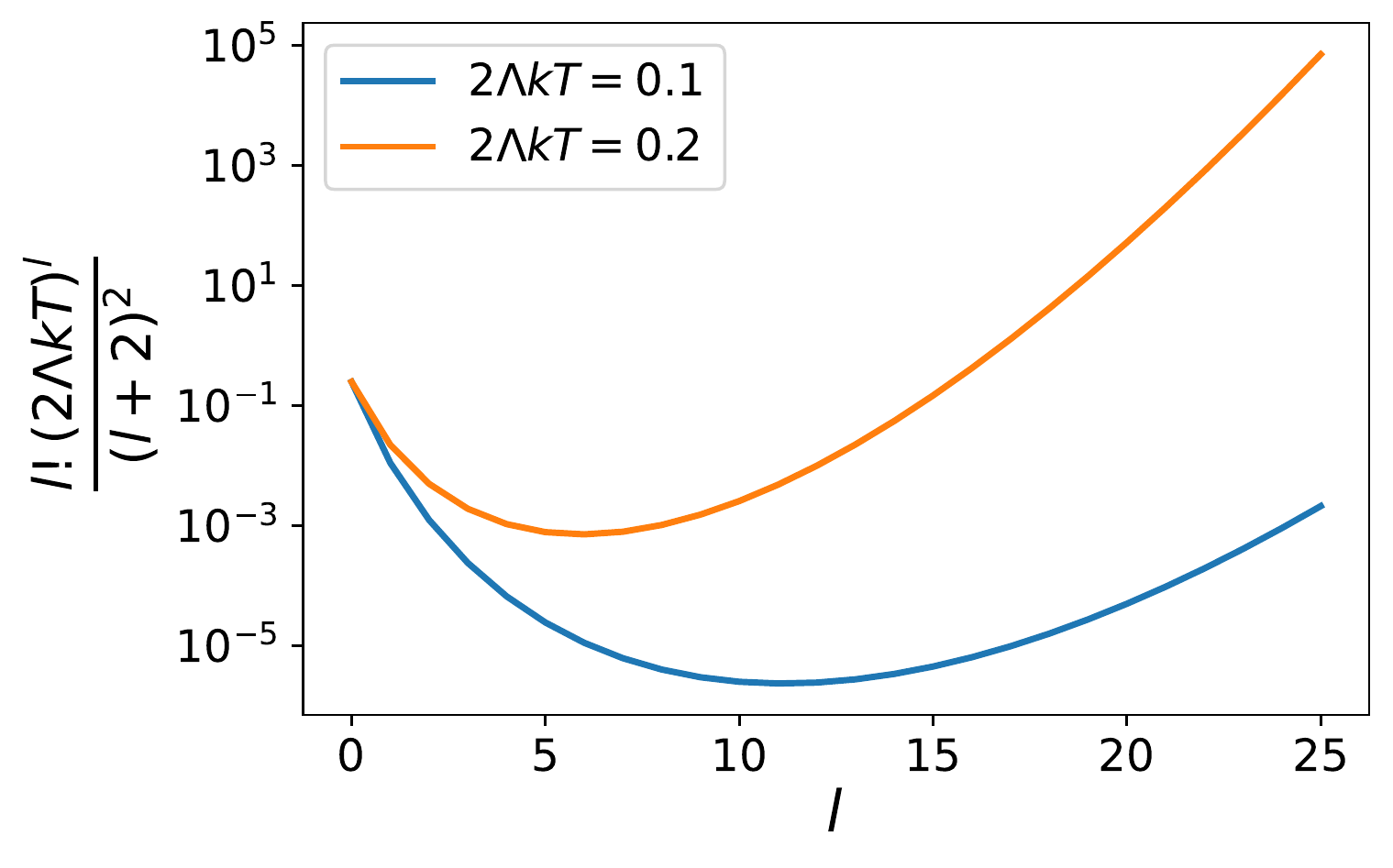}
    \caption{Plot of the upper bound of $T^l\|\hat{H}_l\|/(2N\Lambda)$ given in \cref{eq:upper_FM}. For small $l$, the expansion looks convergent, but when $l$ exceeds $n^*$, where $n^*\propto 1/T\sim\Omega$, it begins to diverge. At the optimal truncation order $l=n^*$, the upper bound behaves as $e^{-\mathcal{O}(\Omega/\Lambda)}$.}
    \label{fig:FM}
\end{figure}

Refs.~\mycite{kuwahara16floquet, mori2016rigorous} obtained a rigorous bound on $H_l$ by noting that, for local Hamiltonians,
\begin{align}
    \|[\hat{H}(t_1),[\hat{H}(t_2),\dots,[\hat{H}(t_l),\hat{H}(t_{l+1})]\dots]]\|
    \leq N\Lambda l!(2\Lambda k)^l.
    \label{eq:comm_bound}
\end{align}
Remarkably, this upper bound is proportional to $N$ for any $l$, and hence the FM expansion ensures extensivity order by order.
Using eq.~\ref{eq:comm_bound}, it was shown that each term of the FM expansion is bounded as
\begin{align}
    \|\hat{H}_l\|T^l\leq 2N\Lambda\frac{l!(2\Lambda kT)^l}{(l+2)^2}.
    \label{eq:upper_FM}
\end{align}
The behavior of this upper bound is illustrated in Fig.~\ref{fig:FM} and matches the results obtained by the numerical solution of microscopic models \cite{d14long}.
It decreases up to $l=n^*\propto\Omega/\Lambda$, but turns to increase for $l>n^*$.
At the optimal order $l=n^*$, the upper bound of $\|\hat{H}_l\|T^l$ is exponentially small: $\|\hat{H}_{n^*}\|T^{n^*}/(N\Lambda)\leq e^{-\mathcal{O}(\Omega/\Lambda)}$. It would then be expected that one can achieve the best approximation by truncating the series at the order $n^*$ and the truncation error is exponentially small in some sense.
Rigorous analyses in ref.~\mycite{kuwahara16floquet, mori2016rigorous} revealed that the precise statement is expressed by \cref{thm:quasi,thm:dyn} with $H_\mathrm{eff}=\sum_{l=0}^{n^*}H_lT^l$.

This rigorous bound can be also understood using the perturbative argument described in the introduction: As  mentioned, $\hat{H}_l$ contains nested commutators like $[\hat{H}(t_1),[\hat{H}(t_2),\dots,[\hat{H}(t_l),\hat{H}(t_{l+1})]\dots]]$.
Because $\hat{H}(t)$ has up to $k$-body interactions, $\hat{H}_l$ contains at most $k(l+1)$-body interaction terms.
Since the local energy scale is bounded by $\Lambda$, the amount of the energy change by applying $\hat{H}_l$ is at most $\Delta E_l=\Lambda k(l+1)$.
If it does not exceed a single energy quantum $\Omega$, the system cannot absorb the energy from the driving field through the $l$th-order term of the FM expansion.
This means that the $l$th-order term of the FM expansion with $l\lesssim \Omega/(\Lambda k)\sim n^*$ is not responsible for heating.
We therefore expect that a truncation of the FM expansion at the $n$th order with $n\leq n^*$ gives an effective Hamiltonian before heating takes place.
This is an intuitive reason why $\hat{H}_{\rm eff}$ describes nonequilibrium dynamics up to the prethermal regime.
On the other hand, the diverging behavior of the FM expansion for $l>n^*$ reflects the fact that heating is a nonperturbative phenomenon in $T=2\pi/\Omega$.

\subsection{Rigorous theory based on renormalization of time-dependent Hamiltonian}
\label{sec:renormalization}


We now move to a   complementary but related   approach, introduced   by refs.~\mycite{abanin2017rigorous,abanin17effective}.
Instead of considering the Floquet unitary which generates stroboscopic dynamics, the focus is shifted onto the  {\it Hamiltonian} which generates the dynamics. 

A key observation is that     dynamics of any given   system can be viewed in different  rotating frames   of reference, such that the   same dynamics can be described by the composition of  (i) motion {\it along} a given frame (as dictated by the frame change), and  (ii) motion {\it within} the frame, the latter of which is generated by some potentially time-dependent Hamiltonian. Concretely, the Hamiltonian in frame $a$ is related to the Hamiltonian in another frame $b$ via the gauge transformation $H_b(t) = Q(t)^\dagger (H_a(t) - i \partial_t) Q(t)$, where $Q(t)$ is the unitary frame change from $a$ to $b$, so that the unitary time-evolution operator can be decomposed as
\begin{align}
    U(t) \equiv \mathcal{T}e^{-i \int_0^t H_a(t')} = Q(t) \mathcal{T}e^{-i \int_0^t H_b(t') dt'}.
    \label{eqn:frame_change}
\end{align}
%
The main idea of refs.~\mycite{abanin2017rigorous,abanin17effective} is that under certain scenarios, such as in the case of a quantum system driven at high frequencies, the frame change $Q(t)$ can be astutely chosen 
such that it is regular (so that its effects at late times can be safely ignored) and that the new Hamiltonian $H_b(t)$    becomes less time-dependent than the previous one, $H_a(t)$.
Such a transformation is often referred to as integrating out the fast degrees of freedom, and leaves behind a renormalized, coarse-grained action on the slow degrees of freedom.
%
%
One can then attempt to repeatedly find additional frame changes to systematically reduce the resulting Hamiltonian's time-dependence; if this can be  eliminated, then heating would be completely arrested, since when viewed in the reference frame resulting from the amalgamation of all the frame changes $Q(t)$, there is a static Hamiltonian governing dynamics.
Of course, as mentioned, for a many-body system, it should not be expected that this procedure can be carried out ad infinitum, and there should be some optimal order to halt the renormalization procedure, such that {\it slow heating} is instead realized.

%
%

 The above intuition was made rigorous in refs.~\mycite{abanin2017rigorous,abanin17effective} for the case of quantum many-body systems with (geometrically) local interactions.
 Their analysis relies importantly on the Lieb-Robinson bound\cite{lieb1972finite}, which states that the velocity of propagation of local correlations is bounded by the local norm  $\Lambda$ of the system (see refs.~\mycite{abanin2017rigorous,abanin17effective} for the precise definition of $\Lambda$ used). Using this, one can estimate very generally that the error in the Heisenberg dynamics $O_i(t)$ of an operator $O$ initially supported  in some local region of space, time-evolved by two systems $i = 1,2$ differing only by an extensive but local quantum many-body Hamiltonian $V(t)$  with local amplitude $J_V$, can be bounded as:
 \begin{align}
     \| O_1(t) - O_2(t) \| \leq C  J_V t (\Lambda t + C')^{d},
     \label{eqn:diff_obs}
 \end{align}
 where $C,C'$ are constants depending only on the choice of operator $O$, and $d$ is the spatial dimension of the system. The above bound simply expresses the   fact that as far as  local observables go, an extensive but weak (as measured by the local norm $J_V$) perturbation will  not affect   dynamics for a long time. Note that a similar statement of closeness in time of two quantum many-body states does {\it not} hold due to the Anderson orthogonality catastrophe; it is essential one considers {\it local} operators. 
 
 With this in mind, refs.~\mycite{abanin2017rigorous,abanin17effective} introduced a renormalization procedure to  choose appropriate rotating frames to render the time-periodic Hamiltonian $H(t)$ of a Floquet system as time-independent as possible, as measured by the local norm.
 Here, refs.~\mycite{abanin2017rigorous,abanin17effective} differ  in their specific   approaches technically: ref.~\mycite{abanin2017rigorous} considered a frame change of the form
 \begin{align}
     Q(t) = \exp\left( -i \sum_{n=0}^{n_\text{max}}  A_n(t) \right),
 \end{align}
 while ref.~\mycite{abanin17effective} considered  a frame change of the form
 \begin{align}
     Q(t) = \prod_{n=0}^{n_\text{max}} Q_n(t) =  \prod_{n=1}^{n_\text{max}} e^{-i A_n(t) },
     \label{eqn:Q}
 \end{align}
where  $A_n(t) = A_n(t+T)$ is a local quantum many-body Hamiltonian to be determined. Here, $n$ denotes the renormalization step, and the maximum step $n_\text{max}$ of the series/product is also to be determined.  For brevity of presentation, we concentrate on the latter approach (though the logic involved in the former is identical, albeit with differing technical steps). In this case, a new Hamiltonian $H_{n+1}(t)$ is sequentially derived from the previous one  $H_n(t)$, by moving into the frame defined by $Q_n(t) = e^{-i A_n(t)}$, with $H_0(t) \equiv H(t)$ being the original time-dependent Hamiltonian that one begins with.
Precisely, at each step  the time-independent part of the Hamiltonian $H_n(t)$ can be defined, via time-averaging:
\begin{align}
D_n = \frac{1}{T} \int_0^T H_n(t),
\end{align}
yielding also the time-dependent part
\begin{align}
    V_n(t) = H_n(t) - D_n.
\end{align}
Ref.~\mycite{abanin17effective} picked the generator $A_n(t)$ of the frame change to satisfy
\begin{align}
    V_n(t) - \partial_t A_n(t) = 0, \qquad A_n(0) = 0.
    \label{eqn:An}
\end{align}
The solution $A_n(t)$ is an extensive operator which has a local norm reduced with respect to $V_n(t)$ by a factor of the inverse frequency $1/\Omega$, and is time-periodic. 
Then, the Hamiltonian at the next step is
\begin{align}
    H_{n+1}(t) = e^{-i A_n(t)} \left(H_n(t) - i \partial_t \right)e^{i A_n(t)},\label{eq:HN1}
\end{align}
which can be immediately seen to again be a local quantum many-body Hamiltonian.

 To see why the choice of frame of eq.~\eqref{eqn:An} is useful, upon splitting $H_{n+1}(t)$ into its time-independent part $D_{n+1}$ and time-dependent part $V_{n+1}(t)$, one may express
 \begin{align}
     V_{n+1}(t) &= \left( \gamma_n(D_n) - D_n \right) + \left( \gamma_n(V_n) - V_n \right) \nonumber \\
     & - \left(\alpha_n(V_n) - V_n\right),
     \label{eqn:Vn}
 \end{align}
 where $\gamma_n(O) = e^{-i A_n(t)} O e^{i A_n(t)}$ and $\alpha_n(O) = \int_0^1 ds e^{-i s A_n(t)} O e^{i s A_n(t)}$. Since $A_n(t)$ is $1/\Omega$ smaller than $V_n(t)$, eq.~\eqref{eqn:Vn} expresses the fact that $V_{n+1}(t)$ is also roughly speaking, $1/\Omega$ smaller than $V_n(t)$ (one can see this by  expanding $e^X Y e^{-X} = Y + [X,Y]+\frac{1}{2}[X,[X,Y]] + \cdots)$. That is, the time-dependence of the renormalized Hamiltonian $H_{n+1}(t)$ has been reduced with respect to that of $H_n(t)$, accomplishing the desired renormalization. In practice, this estimate is not entirely correct: because of the multiple nested commutators in its definition, $V_{n+1}(t)$ is also becoming a more non-local operator, so that its local norm is actually not a monotonically decreasing function of $n$. This behavior is analogous to the non-monotonous behavior of the FM series shown in Fig.~\ref{fig:FM}. 
 A careful estimate yields that there is an optimal order $n_\text{max} = n^* \propto \Omega/\Lambda$ to stop the renormalization procedure at\cite{abanin17effective, PhysRevX.10.021032}, whereupon the local norm of $V_{n^*}(t) $ is estimated to be $ \sim c^{n^*+1}$ for some system-independent constant $c < 1$, i.e., the residual time-dependence is {\it exponentially weak} in the driving frequency. Thus, up to the net frame change $Q(t)$, the time-independent Hamiltonian $D_{n^*}$ captures well the dynamics of local observables up to times which are exponentially long in the driving frequency, see eq.~\eqref{eqn:diff_obs} with $J_V \sim c^{-(n_\text{max}+1)}$. This statement is equivalent to \cref{thm:dyn} with $H_{\rm eff} =  D_{n^*}$.

\section{From quantum spin systems to classical models}

\label{sec:quantum2classical}

The rigorous theorems described in the previous section are directly applicable to quantum many-body systems with local bounded Hilbert spaces. Interestingly, these theorems can be extended to situations where the local Hilbert space dimension can be allowed to grow, as long as the local frequency upper bound $\Lambda$  remains finite. This observation offers a pathway to extend these rigorous theorems to the classical world \cite{mori2018floquet}, which we now review. In particular, one can consider quantum spin chains, with increasing sizes of the local spins, $S$. When $S\to\infty$ the expectation values of the spin components are well described by classical equations of motion \footnote{See ref.~\protect\mycite{assabook} for an introduction.}. In accordance with theoretical arguments provided below, numerical studies of one dimensional chains of classical spins have indeed observed the exponential suppression of the heating rate at large driving frequencies \cite{howell2019asymptotic}. 

We now explain in more detail the connection between classical and quantum spin systems.
To be specific,  consider the following one-dimensional quantum spin-$S$ Hamiltonian:
\begin{align}
    H(t)=&-\frac{1}{2S}\sum_{i,j=1}^N\sum_{\alpha,\beta=x,y,z}J_{ij}^{\alpha\beta}(t)S_i^\alpha S_j^\beta \nonumber \\
    &-\sum_{i=1}^N\sum_{\alpha=x,y,z}h_i^\alpha(t)S_i^\alpha,
    \label{eq:spin-S}
\end{align}
where $J_{ij}^{\alpha\beta}(t)=J_{ij}^{\alpha\beta}(t+T)$ and $h_i^\alpha(t)=h_i^\alpha(t+T)$ are two-body interactions and local magnetic fields, respectively, and $S_i^\alpha$ denotes the $\alpha$-component of a spin-$S$ operator satisfying $\sum_{\alpha=x,y,z}(S_i^\alpha)^2=S(S+1)$.
We note that \cref{eq:spin-S} is a 2-local Hamiltonian.

In the limit of $S\to\infty$, the quantum dynamics is reduced to the classical one.
Let us consider a factorized initial state $\ket{\Psi(0)}=\otimes_{i=1}^N\ket{\psi_i(0)}$, where $\ket{\psi_i(0)}$ is a state vector of $i$th spin.
It is also assumed that the initial state is classical in the sense that
\begin{align}
    \sum_{\alpha=x,y,z}\braket{\psi_i(0)|S_i^\alpha|\psi_i(0)}^2=S^2.
\end{align}
The many-body quantum state $\ket{\Psi(t)}$ evolves under the Schr\"odinger equation $i\partial_t\ket{\Psi(t)}=H(t)\ket{\Psi(t)}$.
Under this setting, normalized spin vectors $\bm{s}_i(t)=(s_i^x(t),s_i^y(t),s_i^z(t))$ with $s_i^\alpha(t)=\braket{\Psi(t)|S_i^\alpha|\Psi(t)}/S$ obey the classical equations of motion\cite{mori2018floquet}, $\partial_t\bm{s}_i(t)=\bm{s}_i(t)\times\tilde{\bm{h}}_i(t)$, where $\tilde{\bm{h}}_i(t)=(\tilde{h}_i^x(t),\tilde{h}_i^y(t),\tilde{h}_i^z(t))$ is the local effective field at site $i$ that is given by $\tilde{h}_i^\alpha=h_i^\alpha+\sum_{j=1}^N\sum_{\beta=x,y,z}J_{ij}^{\alpha\beta}s_j^\beta(t)$.
This is nothing but the classical dynamics with the Hamiltonian $H^\mathrm{cl}(t)=-(1/2)\sum_{i,j}\sum_{\alpha,\beta}J_{ij}^{\alpha\beta}(t)s_i^\alpha s_j^\beta-\sum_{i,\alpha}h_i^\alpha(t)s_i^\alpha$.
Moreover, any quantum correlation function is reduced to the corresponding product of classical spins:
\begin{align}
    &\lim_{S\to\infty}\frac{1}{S^n}\braket{\Psi(t)|S_{i_1}^{\alpha_1}S_{i_2}^{\alpha_2}\dots S_{i_n}^{\alpha_n}|\Psi(t)}\nonumber \\&
    =s_{i_1}^{\alpha_1}(t)s_{i_2}^{\alpha_2}(t)\dots s_{i_n}^{\alpha_n}(t).
\end{align}
The above discussion suggests that one can investigate periodically driven classical spins by considering the infinite-$S$ limit of quantum dynamics instead of directly tackling classical equations of motion.

The rigorous results for quantum spin systems discussed in the previous section cannot be immediately applied to the Hamiltonian in \cref{eq:spin-S}, since the local norm $\Lambda$ is unbounded in the classical limit $\Lambda\propto S\to\infty$. Ref.~\mycite{mori2018floquet} proposed a reformulation of the problem which circumvents this difficulty: each spin-$S$ operator $S_i^\alpha$ is decomposed into $2S$ spin-1/2 Pauli operators $\{\sigma_{i,a}^\alpha\}$ as follows,
\begin{align}
    S_i^\alpha=\frac{1}{2}\sum_{a=1}^{2S}\sigma_{i,a}^\alpha.
\end{align}
Eq.~\ref{eq:spin-S} then assumes the following form, 
\begin{align}
    H(t)&=-\frac{1}{8S}\sum_{(i,a),(j,b)}\sum_{\alpha,\beta}J_{ij}^{\alpha\beta}(t)\sigma_{i,a}^\alpha\sigma_{j,b}^\beta \nonumber \\
    &-\frac{1}{2}\sum_{(i,a)}\sum_\alpha h_i^\alpha(t)\sigma_{i,a}^\alpha.
    \label{eq:spin-S2}
\end{align}
Viewing $(i,a)$ as a site in a \textit{two-dimensional} lattice, 
we find that the \cref{eq:spin-S2} is still 2-local, and its local norm is now finite even in the classical limit. Therefore, \cref{thm:quasi} holds for \cref{eq:spin-S2} in the limit of $S\to\infty$, which proves exponentially slow heating in classical spin systems.

On the other hand, \cref{thm:dyn} cannot be readily extended to classical spins. Indeed,  
this theorem requires geometrically short-ranged interactions, while the Hamiltonian in \cref{eq:spin-S2} describes spins-1/2 on a 2d lattice with non-local interactions (along one spatial direction), even though the corresponding classical Hamiltonian $H^\mathrm{cl}(t)$ is short-ranged. 
Reference \mycite{mori2018floquet} pointed out that a weaker version of \cref{thm:dyn} still holds: the effective Hamiltonian generates an approximate dynamics of classical spin variables up to a time \textit{proportional to} $\omega$, as opposed to the exponentially long heating time. Hence, somewhat counter-intuitively, $H_\mathrm{eff}$ may not describe prethermal dynamics of generic local observables at times where $H_{\rm eff}$ is approximately conserved. Physically, the reason why \cref{thm:dyn} no longer holds is because of chaoticity of   classical dynamics in general: an approximation error $\epsilon(t)$ due to the use of $H_\mathrm{eff}$ is initially exponentially small in $\omega$ but grows exponentially fast due to the butterfly effect, $\epsilon(t)\sim e^{Ct-C'\omega/\Lambda}$ with $C$ and $C'$ being positive constants (here, $\epsilon(t)$ is, for example, defined as the discrepancy between the exact value of $s_i^\alpha(t)$ and its approximate value generated by the Hamiltonian dynamics under $H_\mathrm{eff}$). Finally, it should be emphasized that the inapplicability of \cref{thm:dyn} for dynamics starting from a given initial state does not necessarily imply that $H_\mathrm{eff}$ cannot be used as a proper description of the prethermal regime: It was pointed out that $H_\mathrm{eff}$ captures the prethermal dynamics of an \textit{ensemble of trajectories}~\citep{ye2021floquet}, rather than a single trajectory. 

\section{Statistical arguments for systems without local bounds}

\label{sec:statistical}

\begin{figure}[t]
    \centering
    \includegraphics[width=\linewidth]{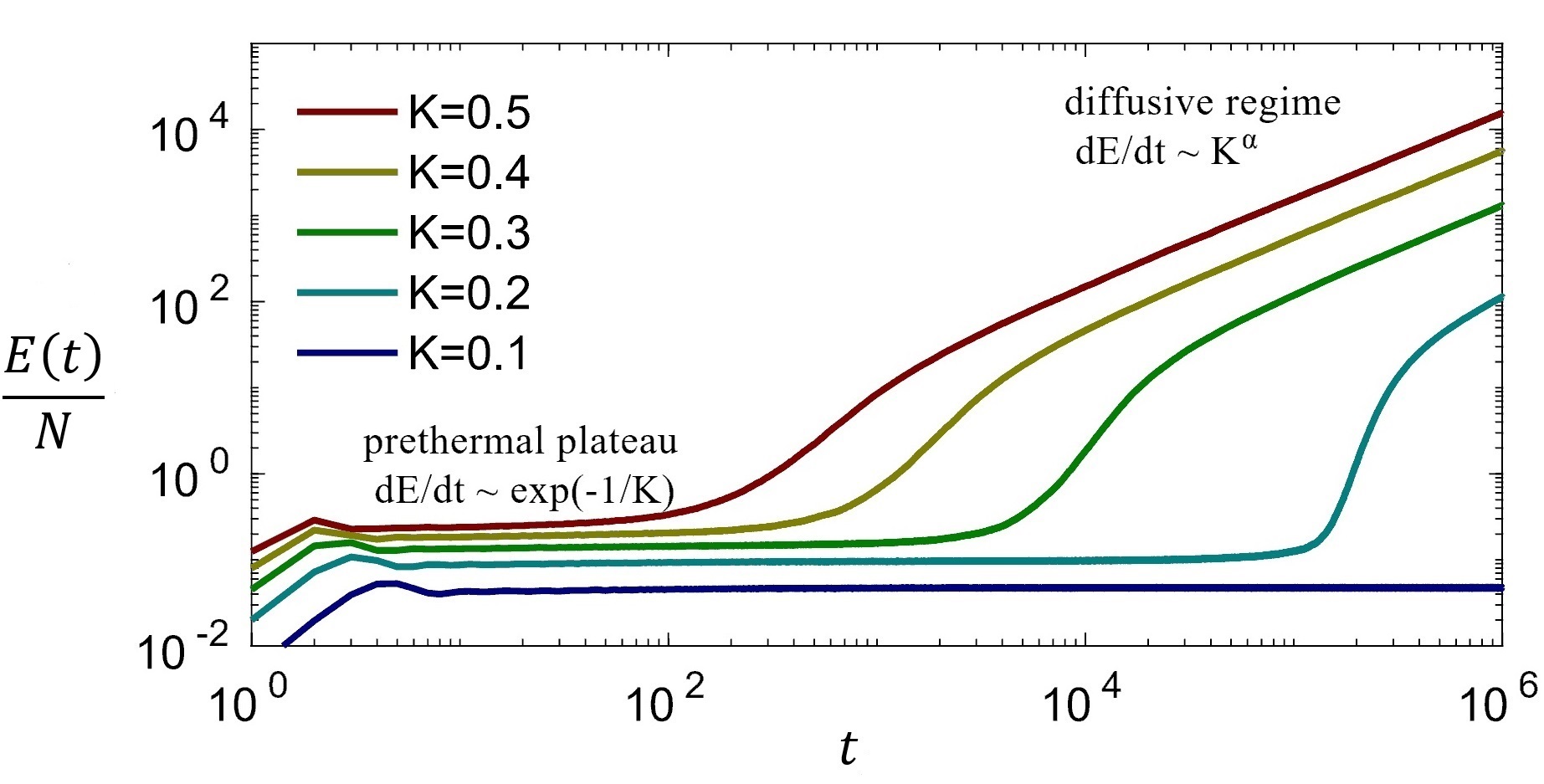}
    \caption{Time evolution of the energy density in a system of coupled kicked rotors, demonstrating the existence of an exponentially long prethermal plateau. The kick strength is set by $K$ and all other parameters are set to unit, $L=T=1$. The plots were obtained the numerical simulation of a chain of 400 rotors and adapted from ref.~\mycite{rajak2019characterizations}. }
    \label{fig:rajak2019}
\end{figure}

We now move to physical systems which do not satisfy the conditions required by the rigorous theorems. A primary example is provided by systems of interacting particles in free space, whose kinetic energy, being quadratic in momentum, is an unbounded operator. Two other important examples of unbounded local Hamiltonians are interacting bosons on a lattice where the potential energy is proportional to the square of the number of bosons per site, and rotors whose kinetic energy is proportional to the square of the angular velocity. In these systems the Hamiltonian includes a quadratic term that can be arbitrarily large and does not have a local bound. Upon driving such systems, it is not clear that an exponentially large suppression of the heating rate is to be expected at high frequencies \cite{hodson2021energy}.
Indeed, earlier numerical studies of periodically kicked classical rotors \footnote{These models are many-body generalizations of the celebrated Chirikov map, one of the best-studied cases of a transition between regular and chaotic motion, see ref.~\protect\mycite{chirikov2008chirikov} for an introduction.} found that the heating rate depends   polynomially on the ratio between the kick strength and its frequency \cite{kaneko89diffusion,konishi90diffusion,falcioni91ergodic,mulansky11strong}. This effect, dubbed ``fast Arnold diffusion'', was explained using tools of many-body chaos 
\cite{chirikov1993theory,chirikov97arnold}. 
Importantly, these studies focused on asymptotically long times only and disregarded the possibility that the system could show interesting transient dynamics. Such intermediate regime was addressed by ref.~\mycite{rajak2018stability}, who studied the transient dynamics of coupled kicked rotors and found numerical evidence of an exponentially long {\it prethermal} plateau, see Fig.~\ref{fig:rajak2019}. 
As explained by ref.~\mycite{rajak2019characterizations}, the suppression of heating in this regime has a {\it statistical} origin, which we now present schematically.

Consider an interacting system that is initially prepared in a low-temperature state. Next, a periodic kick is turned on at a large frequencies, leading to energy absorption. According to the theory of many-body resonances \cite{chirikov79universal}, the heating process occurs through  transitions at resonances, i.e., natural oscillations of the system that have the same frequency as the drive, or its integer multiples. If the frequency is large and the system's spectrum is bounded, such resonances are excluded at low orders in the drive strength, and the heating can be strongly suppressed. In contrast, if the system's spectrum is unbounded, many-body resonances can be found at arbitrarily large frequencies. To compute the heating rate of the system, one has to estimate the probability to encounter such a resonance. As we will see, for (pre)thermal states governed by the Boltzmann distribution, this probability is an exponential function of the driving frequency, leading to an exponential suppression of the heating rate.


To quantify this effect, refs.~\mycite{rajak2018stability,rajak2019characterizations,sadia2021prethermalization} considered coupled kicked rotors, described by the Hamiltonian
\begin{align}\label{eq:rotors}
    H&=\sum_i \frac{1}2 L p_i^2 - K \Delta(t) \sum_i \cos(\phi_i-\phi_{i+1}),\\
    &~~~{\rm with}~~~\Delta(t)=\sum_n \nonumber \delta(t-nT).
\end{align}
Here,  $L$ is the rotor's moment of intertia, $p_i$ is the angular momentum, while the term proportional to the parameter $K$ describes a  periodically modulated interaction. A many-body resonance occurs when a linear combination of frequencies of natural oscillations matches a multiple of the driving frequency $\Omega=2\pi/T$. Treating the interaction term as a perturbation, such that the natural oscillation frequency of the $i$-th rotor is $Lp_i$, this yields the following resonance condition: 
\begin{align}
L(p_i-p_{i+1})=m\Omega. 
\end{align}
with integer $m=\pm1,\pm2, ...$. Hence, in order to hit a resonance, the angular momentum of one of the rotors (or both) must be comparable to the driving frequency, in the appropriate units. Accordingly, the heating rate is roughly proportional to the probability of finding a rotor at $p_i=m\Omega/L$. In the prethermal regime, the distribution function of $p_i$ is given by the Boltzmann distribution, eq.~(\ref{eq:boltzmann}). According to the virial theorem, the angular momenta are not correlated to the angle variables, and one can derive the distribution of the former using eq.~(\ref{eq:boltzmann}) with $H_{\rm eff}= \sum_i L p_i^2/2$ and an inverse temperature $\beta_{\rm eff}$ determined by the energy of the initial state. Thus, the heating rate is expected to be proportional to $\exp(-\beta_{\rm eff} \Omega^2/2L)$, where we considered the lowest order resonance ($m=\pm 1$) and neglected the contributions from $m>1$. For a fixed $\beta_{\rm eff}$ the heating rate is exponentially suppressed at large $\Omega$, giving rise to the phenomenon of statistical Floquet prethermalization\footnote{In the specific case considered by refs. \protect\onlinecite{rajak2018stability,rajak2019characterizations}, the initial conditions where $\phi_i\approx p_i=0$, such that $\langle H_0\rangle =(1/T)\protect\int_0^T dt \langle H(T)\rangle =-K/T$. At equilibrium, the energy is evenly distributed between kinetic and potential energy, leading to $L\langle p_i^2\rangle = K/T=\Omega K /2\pi$, which is obtained for $\beta_{\rm eff}L=2\pi/K\Omega$. Under these circumstances, the exponential suppression goes as $\exp(-\pi\Omega/LK)$ and, in analogy to the rigorous case is an exponential function of $\Omega$, rather than a Gaussian}. Unlike the rigorous case, statistical prethermalization occurs only if the initial state has a low initial temperature, such that $\beta_{\rm eff}\Omega^2/2L\gg 1$ \cite{sadia2021prethermalization}. Furthermore, the suppression of heating  does not last forever: After an exponentially long time, the temperature of the system will become comparable to the driving frequency and the heating rate will no longer be suppressed. At the transition between the two regimes, the system can show anomalous diffusion \cite{rajak2018stability,rajak2020stability} and unconventional correlations \cite{kundu2021dynamics}.

\section{Theoretical extensions and applications}
\label{sec:extensions}

Recent studies have explored Floquet prethermalization at the frontiers of applicability of the rigorous theorems. In particular, Refs.~\mycite{kuwahara16floquet,ho_lr_18,machado2019exponentially, Tran19,machado2020long} investigated prethermalization in systems with power-law interactions with variable exponent $\alpha$, which arises in various quantum simulation platforms, e.g.~arrays of Rydberg atoms which interact via van der Waals interactions, and ensembles of nitrogen-vancancy (NV) centers which interact via dipolar interactions. It was pointed out early on~\cite{mori2016rigorous} that slow heating already occurs for $\alpha>D$, where $D$ is the spatial dimension of the system. However, the stronger result, that the effective Hamiltonian also approximately generates evolution of observables for an exponentially long time, was only possible to be shown for for\citep{kuwahara16floquet} $\alpha > 2D$. The gap in the two results appear to originate from a lack of a generalized Lieb-Robinson bound  that is strong enough to yield an algebraic light-cone in the regime $D < \alpha < 2D$. This was further studied in Ref.~\mycite{Tran19} which derived tighter generalized Lieb-Robinson bounds for the case of power-law interactions (see also Refs.~\mycite{PhysRevA.101.022333, PhysRevX.9.031006, PhysRevLett.113.030602}), and therefore provided improved bounds for prethermalization times using the approach of Ref.~\mycite{abanin17effective,abanin2017rigorous}.

Further, Ref.~\mycite{ho_lr_18} considered periodically-driven systems with power-law decaying interactions whose amplitudes are random and sign-changing. This is relevant, for example, for ensembles of nitrogen-vacancy centers with random positions in space, which interact via dipolar interactions and whose sign hence depends on the relative orientation of two spins. It was shown that the disorder-averaged heating rate is exponentially suppressed with frequency for $\alpha > D/2$, where $D$ is the spatial dimension, due to the cancellation of many terms in the dissipative part of the linear response function coming from the randomness. Interestingly,  a prethermal regime can also occur  at small driving frequencies for sufficiently long-ranged systems, due to the fragmentation of the many-body spectrum of the non-driven system into bands \cite{bhakuni2021suppression}. 

Another scenario departing from the strictly time-periodic Floquet driving involves time quasi-periodic drives, generated by two or more (but still a finite number of) driving frequencies $\Omega_n$ which are rationally independent, that is, $\sum_n q_n \Omega_n \neq 0$ for any $q_n \in \mathbb{Q}$.
An example of such a drive is that resulting from a time series obtained from the Fibonacci word \cite{PhysRevX.10.021032}. As the system is not strictly time-periodic,  there does not  exist a Floquet unitary and hence a Floquet Hamiltonian. Thus  the approach of Sec.~\ref{sec:FM} does not generalize; however the renormalization approach of Sec.~\ref{sec:renormalization} which involves renormalizing the generator of time dynamics does, implying that it might still be possible to capture the system's dynamics within a high-frequency expansion.
Ref.~\mycite{PhysRevX.10.021032} explicitly demonstrated  this,
emphasizing the importance of the {\it smoothness} of the drive profile in controlling the expansion and hence determining the heating rate. They rigorously showed that for sufficiently smooth drives, heating is sub-exponentially suppressed in the magnitude of the driving frequency $|\vec{\Omega}|$ for almost all choices of the vector $\vec{\Omega} = (\Omega_1, \Omega_2, \cdots)$, with an exponent  equal to the number of independent drives (which in particular reduces to a pure exponential in the case where there is only one drive, i.e., a Floquet system). 
The physical reason why smoothness is important can be understood in a perturbative fashion:  the quantity $E_{\vec{a}} = \sum_n a_n \Omega_n$ denotes the possible energy that can be absorbed from the drive, where $\vec{a} = (a_1,a_2,\cdots)$ is a list of integers $a_n$ which denotes the number of energy quanta $\Omega_n$ of drive $n$, and which can become arbitrarily small (i.e., resonant) for large $|\vec{a}|$. 
In a smooth drive, such large $a_n$ are naturally absent at low orders and a resonant $E_{\vec{a}}$ requires a high order process  to be realized. Additionally, the same argument shows that the number-theoretic properties of the frequency vector $\vec\Omega$,  like whether it obeys a Diophantine condition or not,   is an important ingredient, as it determines how fast $E_{\vec{a}}$ approaches 0 for large $|\vec{a}|$. Note that such considerations do not occur at all for periodic Floquet systems. For quasiperiodic drives that are not smooth, like the Fibonacci \cite{dumitrescu2018logarithmically} or the Thue-Morse \cite{zhao2021random} step drive, numerical calculations have  demonstrated that  the system still shows a long lived prethermal behavior. Rigorous theorems have demonstrated that in systems under Thue-Morse quasiperiodic drives the heating rate does not depend exponentially (or sub-exponentially) on the driving frequency, but is rather like $(\omega/\Lambda)^{-C\ln(\omega/\Lambda)}$ with a positive constant $C$ \cite{mori2021rigorous}. 

From a practical perspective, Floquet prethermalization is a useful phenomenon
as it allows for the desirable effects of a periodic drive (like the realization of new effective couplings) to manifest themselves, while suppressing the undesirable effects of heating to infinite temperature, for very long times. 
%
%
This provides a firm basis for Floquet engineering approach in complex many-body systems. 
Furthermore, genuinely novel non-equilibrium states of matter with no equilibrium counterpart can be stabilized in the prethermal regime of a Floquet system. 

Here we describe very briefly the
phenomena of discrete time crystals (DTC) \footnote{See ref.~\protect\mycite{else2020discrete} and references therein.}. In analogy to regular crystals that spontaneously break the spatial translational invariance, DTCs are idealized states that spontaneously break  time translation symmetry (specifically in this case, {\it discrete} time translation symmetry). 
%
The main idea is to kick a strongly-interacting system repeatedly in a strong manner, such that the system cycles  between $N\geq 2$ different `simple' configurations (for example, for the case of a system of qubits, these could be a particular classical spin configuration  and its spin-flipped partner); as it takes $N$ periods for the system to come back to itself, this is a {\it subharmonic} response. 
A paradigmatic example is given by the spin-1/2 transverse Ising model in 2D, driven periodically by $\pi$-kicks in the transverse field   with period $T$:
\begin{align}
H(t) = J \sum_{\langle i j \rangle} \sigma^z \sigma^z_j + \frac{\pi}{2 T} \sum_{i} \sigma^x_i \sum_n \delta(t-nT).
\end{align}
Above, $\langle \cdot \rangle$ refers to pairs of sites which are nearest-neighbors.

Clearly, in this model, beginning from any simple factorized configuration of spin-up and downs $|\sigma\rangle$ (which is a many-body eigenstate of the Ising term), it transitions to its spin-flipped version $|\bar{\sigma}\rangle$ and back after every Floquet cycle.
Note that written as it is, one cannot directly employ a high-frequency expansion to derive an effective Hamiltonian governing dynamics, as the ratio of driving amplitude to frequency $\sim \frac{\pi}{2T} \times T = \pi/2$ is not small.
Now, upon perturbing the Hamiltonian $H(t) \mapsto H(t) + V(t)$, where $V(t)$ represents additional small but arbitrary interactions  with similar time-periodicity,  one would na\"ively  expect that there will be a loss of contrast of the subharmonic behavior over time, owing to spins becoming  more entangled (due to the perturbation) and hence locally incoherent. 
However, by moving into the rotating frame of the   strong kick and when $JT \ll 1$, the
resulting time-periodic Hamiltonian can now be understood as being driven by a weak, high-frequency drive 
which then falls under the purview of the rigorous theorems of Floquet prethermalization --- one can derive an effective static Hamiltonian which governs dynamics till late times. Furthermore, importantly,   an  emergent  discrete $\mathbb{Z}_2$  symmetry   is {\it guaranteed} to be present in the effective Hamiltonian, namely $[H_\text{eff}, \prod_i \sigma^x_i] = 0$, regardless of the precise form of $V(t)$ (more generally, a $\mathbb{Z}_N$ symmetry).
One can understand this as arising from the fact that the strong $\pi$-kicks `symmetrize' the system with respect to the spin-flip operation, dynamically. 
This emergent symmetry is crucial, as the interacting Hamiltonian $H_\text{eff}$ may exhibit spontaneous symmetry-breaking below a critical equilibrium temperature (for instance,  the Ising Hamiltonian in 2D has has a stable ferromagnetic phase at low but finite temperatures), see ref.~\mycite{else17prethermal}. 
In such a scenario, there are then initial configurations with low enough energy with respect to the effective Hamiltonian that will stay in one symmetry-broken well for an exponentially long time in the rotating frame; moving back to the lab frame, this translates to a cycling between simple configurations that survives for such long times.

Thus, we see that the combination of strong interactions and Floquet driving   establishes not only a subharmonic response, but also crucially a {\it robust} one, which is the hallmark of a DTC.
%
%
%
The nonequilibrium phase of matter described above, more precisely called a prethermal discrete time crystal, was recently observed experimentally using trapped ions in ref.~\mycite{kyprianidis2021observation}. Prethermal DTC have also been considered theoretically at high temperatures \cite{luitz2020prethermalization} and in classical systems \cite{ye2021floquet,pizzi2021classicalA,pizzi2021classicalB}. Even more exotically, a discrete time quasi-crystal resulting from time quasi-periodic drives, in which multiple time-translation symmtries are spontaneously broken, was  discussed in refs.~\mycite{dumitrescu2018logarithmically, PhysRevX.10.021032}.

\begin{figure*}
	\includegraphics[width=0.8\textwidth]{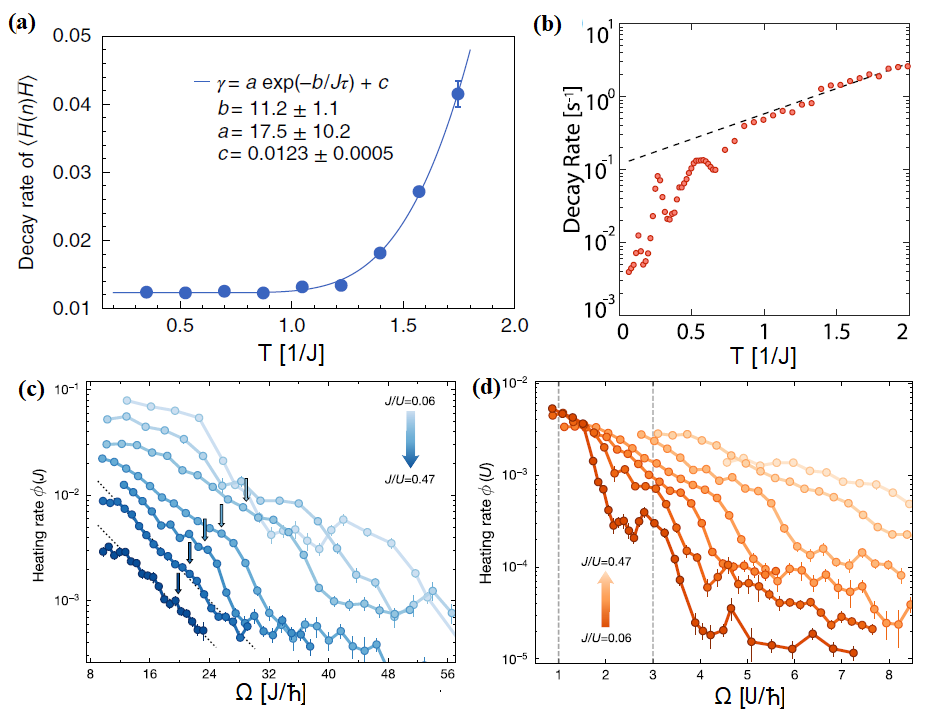}
	\caption{Experimental verification of Floquet prethermalization: heating rates as a function of the drive frequency ($\Omega$) or time period ($T=2\pi/\Omega$) in many-body quantum systems. (a-b) Nuclear spins in (a) flourapatite, reproduced from ref.~\mycite{peng2021floquet}, and (b) in diamond, reproduced from ref.~\mycite{beatrez2021floquet}. Times are measured in units of the spin coupling $J$ and the lines are exponential fits describing Floquet prethermalization. (c-d) Ultracold atoms on an optical lattice with hopping frequency $J$ and on-site interaction $U$: (c) in the Mott insulating regime ($U>J$) and (d) in the superfluid regime ($U<J$), reproduced from ref.~\mycite{rubio2020floquet}. }
	\label{fig:experiments}
\end{figure*}

\section{Experiments}
\label{sec:experiments}

A key assumption in the theory of Floquet prethermalization is that the system is largely isolated from  an external bath and exchanges energy only with the periodic drive. In reality, any physical system is always coupled to some external environment, such as lattice vibration (phonons) or electromagnetic radiation (photons). The resulting  dissipation rate sets a lower limit to the observable heating rate. Hence, in order to observe the predicted  suppression of heating at high frequencies, it is necessary to consider physical systems where heating induced by the environment is several orders of magnitude weaker than the maximal attainable heating rate due to the drive. This requirement is satisfied, for example, in platforms performing analog quantum simulation, 
such as nuclear spins, ultracold atoms,  NV centers, and superconducting circuits. 

Nuclear spins were used in two experiments aimed at probing the rigorous bounds of Floquet prethermalization described in Sec.~\ref{sec:rigorous}. In particular, ref.~\mycite{peng2021floquet} probed the nuclear spins of a mineral called fluorapatite, whose dynamics is approximately described by a one dimensional spin chain model with long coherence time. 
To demonstrate Floquet prethermalization, ref.~\mycite{peng2021floquet} used NMR techniques to probe two-time correlations of the spins. Specifically, the authors focused on temporal correlations of the effective energy operator, $\langle H_{\rm eff}(t)H_{\rm eff}(0)\rangle$, where $\langle \cdot \rangle$ is taken in an infinite-temperature ensemble, see Sec.~\ref{sec:rigorous} for the definition of $H_{\rm eff}$. The decay of this quantity is associated with the heating rate, which can be bounded by rigorous arguments. The experiment confirmed the expected exponential dependence of the heating rates on the drive frequency, see Fig.~\ref{fig:experiments}(a). Interestingly, another approximately conserved quantity of prethermal origin was observed in that experiment. 

Ref.~\mycite{beatrez2021floquet} probed Floquet prethermalization of the nuclear spins of $^{13}C$ isotopes in diamonds. 
The $^{13}C$ nuclei were initially prepared in a coherent state and, in the absence of external pulses, quickly dephase due to large on-site detunings. A periodic drive was used to engineer an effective Hamiltonian that is independent of detuning  and conserves the total magnetization. Further, the decay of the total magnetization was used as a proxy of thermalization. The periodic drive was found to increase the magnetization decay time by more than 5 orders of magnitude. The latter quantity depended exponentially on the ratio between the driving frequency and  interaction energy scale, see Fig.~\ref{fig:experiments}(b), signalling prethermalization.
	
Floquet prethermalization was also probed in systems of ultracold atoms in optical lattices \footnote{The use of ultracold atoms in optical lattices to study Floquet prethermalization was in particular suggested in ref.~\protect\mycite{rajak2019characterizations}.}. In particular, ref.~\mycite{rubio2020floquet} considered optical lattice with approximately one bosonic particle per site, described by the Bose-Hubbard model \cite{bloch2005ultracold}. The system was initially prepared in a low temperature state, and then excited by periodic modulations of the lattice depth. Heating was then probed by measuring the evolution of the number of doubly-occupied sites and of the energy. A prethermal regime with exponentially suppressed heating rates was found. Interestingly, the heating rate exhibited non-monotonic dependence on the particles' interaction strength, attaining a maximum in the proximity of the superfluid-to-Mott insulator quantum phase transition, see Fig.~\ref{fig:experiments}(c-d). Note that the Bose-Hubbard model has an unbounded local Hamiltonian (due to the possibility to pile many bosonic particles on a single site) and in principle cannot be described using rigorous theorems. Nevertheless, the statistical approach reviewed in Sec.~\ref{sec:statistical} could qualitatively capture the non-monotonic behavior of the heating rate, see ref.~\mycite{dalla2021statistical}, wherein two complementary perturbative approaches, valid on the two sides of the transition, were used.

Refs.~\mycite{singh2019quantifying} and \mycite{cao2021interaction} probed the Floquet prethermalization of bosonic atoms in optical lattices at large occupation numbers. In both experiments, the atoms where initially prepared in the ground state, and then excited by a time-periodic change of the optical lattice. Ref.~\mycite{singh2019quantifying} probed heating by measuring the inverse participation rate (IPR) of the system with respect to the eigenstates of the unperturbed Hamiltonian. This quantity equals to $1$ if the system is in its ground state and tends to 0 when a macroscopic number of states is excited. The experiment demonstrated a transition between a regular diffusive regime at low driving frequencies where the $\text{IPR} \sim t^{-1/2}$, to a regime of suppressed heating at large driving frequencies where the IPR $\sim t^{-1/4}$. While a quantitative description of this transition is still lacking, the experiment is in the regime of validity of the semiclassical methods presented in Sec.~\ref{sec:statistical}, where a subdiffusive behavior was numerically observed \cite{rajak2018stability}. In ref.~\mycite{cao2021interaction}, the periodic drive was turned on after having shut down the longitudinal confinement. In the absence of interactions, the motion in the longitudinal direction is described by the celebrated quantum kicked rotor, where the heating rate is completely suppressed due to dynamical localization of the atoms in momentum space  \cite{fishman89scaling}. In the presence of interactions\footnote{Initial numerical findings suggested that the dynamical localization occurs in many-body systems as well \cite{d13many}. These numerical findings were, however, limited to small systems, where it is not possible to distinguish between a parametric reduction of heating induced by Floquet prethermalization and its complete suppression. It was later demonstrated that interactions between the atoms lead to a violation of the dynamical localization and restore the heating \cite{luitz2017absence,notarnicola2018localization}.}, the experiment showed an anomalous diffusion that had been theoretically predicted in ref.~\mycite{notarnicola2018localization}. In spite of the similarity between dynamical localization and Floquet prethermalization, i.e., a strong suppression of heating at large driving frequencies, these two effects have a different physical origin: dynamical localization is due to interference effects that are not necessary for Floquet prethermalization. Integrating these complementary effects  together  requires further theoretical investigations. Finally, ref.~\mycite{shkedrov2022absence} applied a time-periodic drive to ultracold fermions in free space and demonstrated a strong suppression of heating at large driving frequencies.

Additional insights into the Floquet prethermalization were provided by extensive numerical calculations, some of which were performed on high-performance computers. For example, ref.~\mycite{morningstar2022simulation} studied Floquet prethermalization effect using tensor processing units (TPUs), specialized hardware accelerators developed by Google to support large-scale machine-learning tasks. By massively parallelizing the problem over 128 TPUs, they were able to simulate one-dimensional chains with up to 34 spins for $10^5$ Floquet periods. For comparison, Ref.~\mycite{okamoto2021floquet} studied smaller two-dimensional systems with up to $14$ qubits. Refs.~\mycite{rubio2020floquet,dalla2021statistical} simulated the one-dimensional Bose-Hubbard model, with 9 particles on 9 sites. Ref.~\mycite{mallayya2019heating} studied the heating rates of one-dimensional systems in the thermodynamic limit, using a numerical linked cluster expansion. All these numerical works found exponential dependence of the heating rate on the driving frequency, over several orders of magnitude.


\section{Summary and discussion}
\label{sec:summary}
In this paper, we have reviewed the phenomenon of Floquet prethermalization, which is the strong parametric
suppression of heating in periodically-driven many-body systems at large driving frequencies, and described several instances where it occurs, as well as its physical origins. We provided an overview of the rigorous theorems underpinning Floquet prethermalization which apply to spin systems, and more generally to lattice systems with locally bounded Hilbert spaces (e.g., interacting fermions).
We further discussed statistical arguments for prethermalization in several cases which are not directly captured by the rigorous results. We also described several recent experiments performed in systems ranging from nuclear spins to ultracold atoms in optical lattices, that reported signatures of Floquet prethermalization. The  applicability of this concept to a broad range of systems shows that prethermalization is a universal phenomenon that does not require fine-tuning.

Ideas and techniques introduced in the context of Floquet prethermalization have also found important applications in seemingly unrelated problems. One application is the emergence of quasi-conserved quantities in static systems with a large separation of energy scales. A concrete example is the dynamics of doublons in the Bose-Hubbard model with strong on-site interactions. Experimentally, the relaxation time of a doublon had previously been found to be exponentially long in the interaction strength~\cite{DoublonDecay}, strongly suggesting that the total number of doublons is a quasi-conserved quantity. Using techniques of the rigorous theorems described in ref.~\mycite{abanin2017rigorous}, this can be  proven. 
Indeed, by moving into a frame of reference where states with different number of doublons oscillate at different frequencies separated by the large interaction strength, 
the system can be mapped to a periodically-driven one wherein terms describing non-doublon conserving transitions oscillate quickly. This then falls under the purview of the rigorous theorems of Floquet prethermalization, allowing one to derive an effective Hamiltonian which captures approximate conservation of energy for an exponentially long time. However, additionally and importantly, because of the special structure of the terms which are driven, it can also be rigorously shown that the effective Hamiltonian lacks any non-doublon conserving terms, since they have been  `integrated out'. That is to say, the effective Hamiltonian harbors an emergent $U(1)$ symmetry pertaining to the doublon number, which is conserved for the same long times. In fact, one can even consider generalizations of the above setup where the large energy separations are themselves periodically-modulated in time.
It was shown in Ref.~\mycite{ho2020rigorous} that intriguingly, by working in the opposite limit wherein these modulations are slowly-varying (i.e., the {\it low-frequency} regime), one can  derive a description of dynamics (in a new frame) such that the generator of dynamics has the aforementioned emergent $U(1)$ symmetry, yet is no longer time-independent. This leads to an interesting and exotic scenario where such a system may quickly heat up (due to a lack of energy conservation), yet conserves the charge associated with the emergent $U(1)$ symmetry for long times --- a phenomenon that has  been dubbed {\it prethermalization without temperature} (see also ref.~\mycite{luitz2020prethermalization}).
These works show that the theoretical methods developed for Floquet prethermalization are very general and have a broad applicability.

Looking ahead, one of the outstanding challenges is to develop new rigorous techniques to establish prethermalization that go beyond the results reviewed in this paper. One avenue where experiments and numerics suggest that there may be room for rigorous results, is that of driven systems of interacting bosons on a lattice, such as the Bose-Hubbard model studied in ref.~\onlinecite{rubio2020floquet}.
These studies also demonstrated exponentially slow heating, yet current rigorous approaches to Floquet prethermalization cannot handle this setup, owing to the unbounded nature of the local Hilbert space. 
Another avenue is in closing the gap between bounds of heating coming from linear-response arguments  and the ability to employ a Magnus-expansion\cite{Tran19} in long-range interacting systems, depending on the power-law exponent, as described in Sec.~\ref{sec:extensions}.  It will also be interesting to investigate whether prethermalization bounds may be strengthened in cases when multiple conservation laws are present in the system. 
On the experimental front, we expect that the Floquet prethermalization ideas will keep finding new applications, providing a useful framework for many-body state preparation and manipulation. One of the interesting open questions in that direction is to describe the effect of a weak external heat-bath on quasi-conserved quantities. Developing a better understanding of this effect may open the door to applications of Floquet prethermalization to engineering collective states of electrons in solid-state materials.

\begin{acknowledgments}
We would also like to thank our collaborators on previous works on this topic, and in particular, Roberta Citro, Itzhack Dana,  
Wojciech De Roeck, Philipp Dumitrescu, Dominic Else, Fran\c{c}ois Huveneers, Tomotaka Kuwahara, Atanu Rajak, and Keiji Saito.
This work was funded by the Israel Science Foundation, Grants No.~151/19 and 154/19 (EGDT). T.~M.~was supported by JSPS KAKENHI Grant No.~JP19K14622 and JP21H05185 and by JST, PRESTO Grant No.~JPMJPR2259. W.~W.~H.~
is supported by the National University of Singapore  start-up grants A-8000599-00-00 and A-8000599-01-00. D.~A.~A.~was supported by the European Research Council (ERC) under the European Union’s Horizon 2020 research and innovation programme (grant agreement No.~864597) and by the Swiss National Science Foundation.
\end{acknowledgments}

\bibliographystyle{naturemag}
\bibliography{references}

\end{document}